\documentclass[aps,pra,showpacs,fleqn,twocolumn]{revtex4}
\usepackage{graphicx}
\usepackage{amsmath}
\usepackage{amsfonts}
\usepackage{amssymb}
\usepackage[svgnames]{xcolor}

\newcommand{\Tr}{\text{Tr}\,}

\newcommand{\RDM}{\rho}

\renewcommand{\imath}{\mathrm{i}}

\newcommand{\be}{\begin{equation}}
\newcommand{\ee}{\end{equation}}
\newcommand{\bdm}{\begin{displaymath}}
\newcommand{\edm}{\end{displaymath}}
\newcommand{\bi}{\begin{itemize}}
\newcommand{\ei}{\end{itemize}}

\renewcommand{\imath}{\mathrm{i}}

\begin{document}
	\title{Entanglement Dynamics of Quantum Oscillators\\ Nonlinearly Coupled to Thermal Environments}
  \author{Aurora Voje}
  \author{Alexander Croy}  
  \author{Andreas Isacsson} \email{andreas.isacsson@chalmers.se} 
  \affiliation{Department of Applied Physics, Chalmers University of Technology, 
  					SE-412 96 G\"{o}teborg, Sweden}

\begin{abstract}
We study the asymptotic entanglement of two quantum harmonic
oscillators nonlinearly coupled to an environment. Coupling to
independent baths and a common bath are investigated. Numerical
results obtained using the Wangsness-Bloch-Redfield method are
supplemented by analytical results in the rotating wave
approximation. The asymptotic negativity as function of temperature,
initial squeezing and coupling strength, is compared to results for
systems with linear system-reservoir coupling. We find that due to the
parity conserving nature of the coupling, the asymptotic entanglement
is considerably more robust than for the linearly damped cases. In
contrast to linearly damped systems, the asymptotic behavior of
entanglement is similar for the two bath configurations in the
nonlinearly damped case. This is due to the two-phonon system-bath
exchange causing a supression of information exchange between the
oscillators via the bath in the common bath configuration at low temperatures.
\end{abstract}
\date{\today}
 %Micro- and nano-electromechanical systems
                %(MEMS/NEMS) and devices
\maketitle

\section{Introduction}
Entanglement challenges our comprehension since the
1930's~\cite{eipo+35,sc35}, and still remains a highly relevant
topic. Problems relating to entanglement creation and manipulation are
of importance for a broad range of questions related to quantum
information science\cite{hoho+09}, like quantum
cryptography\cite{ek91}, quantum dense coding\cite{bewi92}, quantum
computation algorithms \cite{sh95} and quantum state
teleportation~\cite{yust92,bebr+93,bove+98}. In particular, recent
experimental advances \cite{bopa+97,bobr+98,fuso+98,pfhe+14} pave the
way for entanglement-based technology.

In this paper, asymptotic effects of nonlinear dissipation on the
entanglement of harmonic oscillators are investigated and compared to
the widely studied situation of linearly damped (LD) systems.  In the
latter the oscillators are linearly coupled to bosonic
reservoirs. Such system-reservoir interactions have for instance been
investigated within both Markovian \cite{pr04, befl06} and
non-Markovian dissipation models
\cite{maol+07,ligo07,paro08,paro09}. For systems initially in squeezed
states, high temperature entanglement \cite{gapa+10}, and the exotic
behavior of entanglement sudden disappearance and revival (ESDR)
\cite{yueb04,seil+04,pr04,paro08,maol+07} have been found.

Typically dissipation destroys quantum entanglement. However, it is
known that by engineering the system-reservoir coupling, entanglement
can be generated \cite{silo78,krma+13,vocr+13}.  For example, one
possibility to entangle initially separable states is through the
introduction of multi-quanta dissipation, or nonlinear damping
(NLD). Naturally occurring NLD has been reported in systems which
possess strong intrinsic nonlinearities \cite{dykr84}. Among these are
carbon-based nanomechanical systems like graphene and carbon nanotubes
\cite{eimo+11,crmi+12,vocr+13}. Additionally, there have been reports
of inducing nonlinear dissipation in optomechanical systems
\cite{nubo+10,lemi04}, and suggestions for possible emergence of NLD
in solid state quantum devices \cite{evsp+14}.

In Ref.~\cite{vois+13} we demonstrated the possibility of
entanglement generation from initially separable states. Here, in the
light of previous studies on linearly damped oscillator systems, we
investigate how NLD affects the asymptotic state behavior of initially
entangled states.  For the latter we choose two-mode squeezed vacuum
states \cite{pfhe+14,si00,pr04,brva05}.  These states are entangled,
Gaussian states, which approach the maximally entangled EPR-state by
an increase of the squeezing parameter.

While NLD is usually accompanied by a conservative nonlinearity, it
was found in Ref.\ \cite{vois+13} that a weak conservative Duffing
nonlinearity did not affect the asymptotic state behavior when only
the lowest lying eigenstates are occupied. Hence, we here limit the
study to purely harmonic oscillators, nonlinearly coupled to either
one common or two individual environments. This approach serves to
isolate effect of the nonlinear relaxation behavior on the
entanglement and allows a more transparent comparison with the
linearly damped systems.

Compared to a linearly damped system we find that the parity
protection inherent to the two-phonon exchange between the system and
the reservoirs, and present in the nonlinearly damped systems, changes
the asymptotic behavior in several ways.  Firstly, the asymptotic
decay of entanglement is considerably slower for two uncoupled
oscillators. This is due to the necessity of simultaneous excitation
processes of the two oscillators needed for thermal
dephasing. Secondly, for individual oscillators coupled to a common
bath, we do not reproduce the sharp transition between steady state
entanglement and disentanglement in the infinite time limit, seen in
linearly damped systems. For the linearly damped system, persistent
entanglement is connected to the relative oscillator motion degree of
freedom being decoupled from the bath. For the nonlinearly damped
system, no such decoupling occurs. Finally, for weakly coupled
oscillators, parity protection in combination with coherent
oscillations in the oscillator populations due to the coupling leads
to disappearance and reappearance of entanglement reminiscent of ESDR
behavior.

The organization of this paper is: First, in
Sec.~\ref{sec:model}, we present the model Hamiltonians and derive
quantum master equations (QMEs) for two separate harmonic oscillators
coupled to either individual baths or a common bath. Then, in
Sec.~\ref{sec:results}, we present the asymptotic entanglement
behavior as function of temperature, initial squeezing and dissipation
rate. We compare our results on nonlinearly damped systems to
previous results on linearly damped systems. In
Sec.~\ref{sec:coupled}, we comment on some features of the asymptotic
entanglement behavior of a coupled oscillator system.

\section{Quantum master equations for uncoupled oscillators\label{sec:model}}
First we consider two different scenarios where a system of two
independent harmonic oscillators with frequencies $\omega_0$ are
coupled quadratically in position to either two individual, or one
common reservoir of harmonic oscillators. The situation with two
weakly coupled oscillators is discussed in
section~\ref{sec:coupled}. For the uncoupled oscillators the
Hamiltonian is $H=H_{\rm S} + H_{\rm B} + H_{\rm
  SB}^{\rm{cb,ib}}$. Measuring length, time and energy in units of
$\sqrt{\hbar/2m\omega_0}$, $\omega_0^{-1}$ and $\hbar\omega_0$
respectively, we have
\begin{subequations}\label{eq:hamiltonian_terms}
\begin{align}
H_{\rm S} = & \sum_{j=1,2}
\left(\frac{1}{2}p_j^2+\frac{1}{2}\omega_0^2 q_j^2 \right)\;,
\label{eq:hamiltonian_terms1}\\ 
H_{\rm B} = &\sum_j\sum_k \omega_{j k} b^{\dag}_{j k} b_{j
  k}\;,\label{eq:hamiltonian_terms2}\\ H_{\rm SB}^{\rm ib} = &\sum_j
q_j^2 \sum_k \eta_{j k} (b^{\dag}_{jk} +
b_{jk})\;,\label{eq:hamiltonian_terms3}\\ H_{\rm SB}^{\rm cb} =
&\sum_j q_j^2 \sum_k \eta_{k} (b^{\dag}_{k} +
b_{k})\;.\label{eq:hamiltonian_terms4}
\end{align}
\end{subequations}
Here, $p_j= \imath \sqrt{\omega_0/2}(a_j^{\dag}-a_j)$ and
$q_j=(a_j^{\dag}+a_j)/\sqrt{(2 \omega_0)}$ denote momentum and
oscillation amplitude of oscillator $j$, respectively, while $a_j$
$(a_j^{\dag})$ is the annihilation (creation) operator of the $j$-th
oscillator. 

The system-bath coupling part of the total Hamiltonian is denoted by
$H_{\rm SB}^{\rm ib}$ for two individual baths, and $H_{\rm SB}^{\rm
 cb}$, for the common bath. For the individual bath configuration
the operator $b^\dagger_{jk}$ ($b_{jk}$) creates (destroys) a phonon
in state $k$ of reservoir $j$ with the frequency $\omega_{j k}$. The
coupling strength of oscillator $j$ to reservoir state $k$ is denoted
by $\eta_{j k}$. Similarly, for the common bath, the operator
$b^\dagger_{k}$ ($b_{k}$) creates (destroys) a phonon in state $k$ of
the common reservoir with the frequency $\omega_{k}$. The coupling
strengths of both oscillators to the reservoir state $k$ are denoted by
$\eta_{k}$.

To study the time evolution of the system we numerically solve the
QMEs for the reduced density matrix $\RDM$ in the weak
system-reservoir coupling limit. To obtain analytical results we
implement the rotating wave approximation (RWA). 
Below we summarize the QMEs for both bath configurations with
and without RWA.

\subsection{QME for coupling to individual baths}
Using the Born-Markov approximation in the interaction picture with
respect to $H_{\rm S}$, the general QME for the individual bath configuration is
given by \cite{brpe02}
\begin{eqnarray}\label{eq:QMEwIntfinal_ib}
\frac{\partial }{\partial t}\rho(t) =& 
-\sum\limits_{l,j}\int\limits_0^{\infty} d\tau\Big[ 
S_l(t) , S_j(t-\tau)\rho(t)\Big]C_{l j}(\tau) \nonumber\\
&\:-\Big[\:S_l(t),\rho(t) S_j(t-\tau)\Big]
C_{j l}(-\tau)\;.
\end{eqnarray}
The operators $S_j(t) = e^{\imath H_{\rm S} t}\;(a_j^{\dag}
+a_j)^2 \;e^{-\imath H_{\rm S} t}$ and $B_j(t)=
\sum_k{\eta_{jk}}\left(b^{\dag}_{j k} e^{\imath \omega_{j k} t} + b_{j
 k} e^{-\imath \omega_{j k} t}\right)$ 
allow us to rewrite the coupling Hamiltonian as 
$H_{\rm SB}^{\rm ib}(t)= \sum_{j=1,2}
S_j(t)\otimes B_j(t)$ in the interaction picture. Assuming initial
thermal equilibrium of the reservoirs, 
$\rho_{\rm B}=\rho_{{\rm B},1}\otimes\rho_{{\rm B},2}$, their
correlation functions $ C_{jl}(\tau) = \Tr_{\rm \! B} \{B_j(t)
B_l(t-\tau)\rho_{\rm B} \}$ are 
\begin{multline}\label{eq:corrfunct}
	C_{j l}(\tau) = \delta_{j l}
        \int\frac{d\omega}{2\pi}\kappa_{j}(\omega) \left[ N(\omega)
          e^{\imath \omega \tau} \right. \\ \left.+ (N(\omega) + 1)
          e^{-\imath \omega \tau} \right]\;,
\end{multline}
where $N(\omega)=(e^{\omega/k_{\rm B}T}-1)^{-1}$ is the Bose-Einstein
distribution and $\kappa_{j}(\omega)=2\pi \sum_k
|\eta_{jk}|^2\delta(\omega-\omega_{jk})$ are the spectral densities.
The specific form of $\kappa_j$ depends on the microscopic details of
the system-reservoir coupling. If $\kappa_j$ is sufficiently smooth
around the frequencies of interest, the exact frequency dependence is
not crucial. To be specific, we use an Ohmic spectral density,
$\kappa_{j}(\omega)=\Gamma_j \omega/(2\omega_0)$, where $\Gamma_j$ is
the non-linear dissipation strength of the $j$-th bath.

Further, we define the one-sided
Fourier transform of the reservoir correlation function
\begin{equation}\label{eq:gamma_def}
	\frac{1}{2} \gamma_j (\omega)
        +\imath \sigma_j (\omega)\; = \int\limits_0^{\infty} d\tau\; e^{\imath
          \omega \tau} C_{jj}(\tau).
\end{equation}
The rates $\gamma_j$ determine the strength of dissipation, while
$\sigma_j$ renormalize the system Hamiltonian. For simplicity we from
here on let $\omega_0$ denote the renormalized system frequencies and
neglect the corresponding small induced conservative nonlinearity.

Using the expression of the bath correlation function
\eqref{eq:corrfunct} one finds
\begin{subequations}
\begin{align}
	\gamma_j(2\omega_0) ={}&  \Gamma_j [N(2\omega_0) + 1]\;,\\
	\gamma_j(-2\omega_0) ={}& \Gamma_j N(2\omega_0)\;.
\end{align}
\end{subequations}

In the RWA, equation (\ref{eq:QMEwIntfinal_ib}) simplifies to
\begin{eqnarray}\label{eq:QME_RWA_wIntfinal_ib}
\dot{\rho} &= & -\frac{1}{2}\sum_{j=1,2}
\Big[\gamma_j(2\omega_0)\mathcal{L}_1[a_j^{\dag 2}] +\gamma_j(-2\omega_0)\mathcal{L}_1[a_j^2] \Big]\rho,\nonumber\\ &&
\end{eqnarray}
with
\begin{eqnarray} 
\mathcal{L}_1[X_j]\rho &=&  X_j X_j^\dag \rho + \rho X_j X_j^\dag 
-2 X_j^\dag \rho  X_j .
\end{eqnarray}

\subsection{QME for coupling to a common bath}
For the common reservoir configuration the summation in
(\ref{eq:QMEwIntfinal_ib}) can be omitted and the general form of the
common bath QME is
\begin{eqnarray}\label{eq:QMEwIntfinal_cb}
\frac{\partial }{\partial t}\rho(t) =& -\int\limits_0^{\infty}
d\tau\Big[ S(t) , S(t-\tau)\rho(t)\Big]C(\tau)
\nonumber\\ &\:-\Big[\:S(t),\rho(t) S(t-\tau)\Big]C(-\tau),
\end{eqnarray}
where the common system and bath operators in (\ref{eq:QMEwIntfinal_cb}) are
$S(t) = \sum_{j=1,2} (a_j^{\dag}e^{\imath \omega_0 t} + a_je^{-\imath
  \omega_0 t} )^2$ and $B(t)= \sum_k{\eta_{k}}\left(b^{\dag}_{k} e^{\imath
  \omega_{k} t} + b_{k}e^{-\imath \omega_{k} t}\right)$. The correlation 
reservoir function is then given by
$C(\tau)=\Tr_{\rm \! B} \{B(t) B(t-\tau)\rho_{\rm B} \}$.

In this case the interaction picture RWA QME is
\begin{eqnarray}\label{eq:QME_RWA_wIntfinal_cb}
\dot{\rho} &= & -\frac{1}{2}\sum_{j=1,2}
\Big[\gamma_j(2\omega_0)\big(\mathcal{L}_1[a_j^{\dag
      2}]+\mathcal{L}_2[a_j^{\dag 2}]\big)\nonumber\\ &&+
  \gamma_j(-2\omega_0)\big(\mathcal{L}_1[a_j^2]
  +\mathcal{L}_2[a_j^{2}] \big) \Big]\rho
\end{eqnarray}
with
\begin{eqnarray} 
\mathcal{L}_2[X_j]\rho &=&  X_j X_{j-(-1)^j}^\dag \rho + \rho X_j X_{j-(-1)^j}^\dag \\\nonumber
& & -2 X_{j-(-1)^j}^\dag \rho X_j \;.
\end{eqnarray}

The QME in (\ref{eq:QME_RWA_wIntfinal_cb}) is similar to
(\ref{eq:QME_RWA_wIntfinal_ib}), but with additional cross terms by
which the two sub-systems are connected via the bath.

As shown in Ref.~\cite{paro08}, coherence and entanglement of a
squeezed state is better preserved in a symmetric system. We therefore
consider a setup in which the dissipation rates
for the two oscillators are set equal in all system-bath
configurations, $\Gamma_j=\Gamma_0$. We also
define $\gamma(2\omega_0)=\gamma_{2-}$ and
$\gamma(-2\omega_0)=\gamma_{2+}$. Further, we define basis vectors
$\vert n,i \rangle=\vert n\rangle_1 \otimes \vert i\rangle_2$,
denoting eigenstates with $n$ quanta in oscillator 1 and $i$ quanta in
oscillator 2.

\section{Results for independent oscillators\label{sec:results}}
In order to compare the asymptotic entanglement of nonlinearly damped
independent oscillators to linearly damped ones, we solve the QME \eqref{eq:QMEwIntfinal_ib}
numerically. We use the Wangsness-Bloch-Redfield approach in the
eigenbasis of the system Hamiltonian~\cite{wabl53,bl57,re65} in a
Hilbert space truncated above $M=8$ eigenstates for each oscillator.

To facilitate the comparison, the system is initialized with the two-mode
squeezed vacuum 
\be\label{eq:2modesqvac} \rho(0)=\hat{S}_{12}(\xi)
\vert 0,0 \rangle \langle 0,0 \vert \hat{S}^{\dag}_{12}(\xi)\;, 
\ee
where the two-mode squeezing operator is $\hat{S}_{12}(\xi)=e^{\xi
  a^{\dag}_1a^{\dag}_2- \xi^{*} a_1a_2}$ and $\xi=r e^{\imath
  \theta}$. In the Fock basis, using $\theta=\pi$, equation
\eqref{eq:2modesqvac} becomes \cite{drfi04}
\be\label{eq:2modesqvac_fock}
\rho=
\frac{1}{\cosh^2(r)}\sum_{n,m=0}^{\infty} 
(-1)^{(n-m)} [\tanh (r)]^{n+m} \vert n,n\rangle 
\langle m, m \vert.
\ee

To quantify the entanglement we use the measure of negativity
$\mathcal{N}=(||\rho^{{\rm T}_1}||_1 - 1 )/2$, where $\rho^{{\rm
    T}_1}$ denotes the partial transpose of the bipartite density
matrix with respect to oscillator one. The negativity corresponds to the
absolute value of the sum of negative eigenvalues of $\rho^{{\rm
    T}_1}$ and vanishes for separable states~\cite{viwe02}. 
%%%%%%%%%%% FIGURE 1 %%%%%%%%%%%%%%%%%%%%%%%%%%%%%%%%%%%%%%%%%%%%%
\begin{figure}[t]
  \centering
\includegraphics[width=\linewidth]
                      {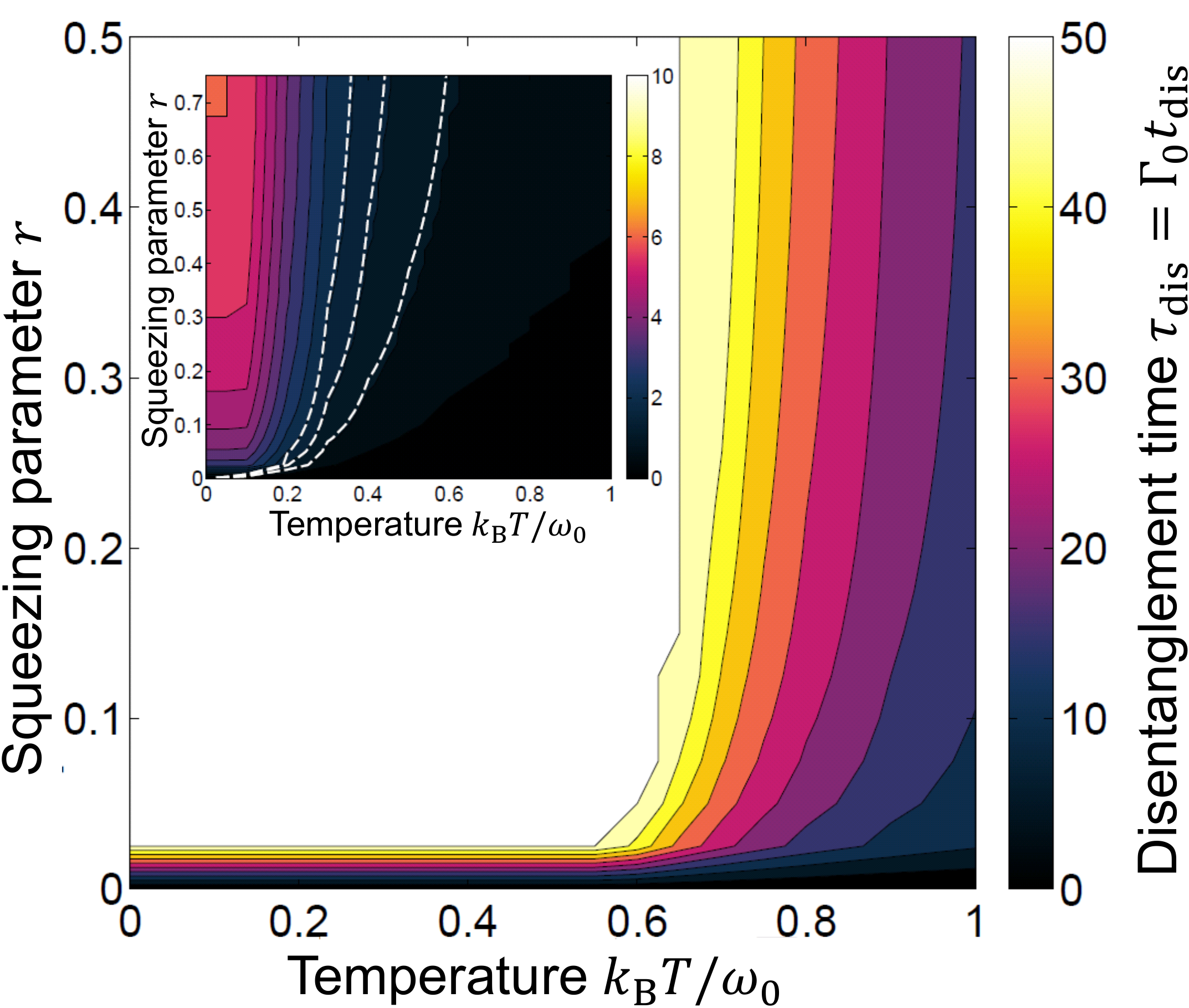}
\caption{(Color online) Main figure: scaled disentanglement time
  $\tau_{\rm dis}=\Gamma_0 t_{\rm dis}$ (colorbar) of nonlinearly
  damped two-mode squeezed vacuum states as function of temperature
  $T$ and squeezing parameter $r$, for two individual baths with
  simulation time $\tau_{\rm sim}=50$, damping rate $\Gamma_{0}=
  10^{-3}\omega_0$, and negativity cut-off $\varepsilon=10^{-3}$.
  Inset: disentanglement time $\tau_{\rm dis}$ (inset colorbar scale)
  of linearly damped two-mode squeezed vacuum states as function of
  $T$ and $r$ for two individual baths, with simulation parameters as
  in the main figure.  The dashed lines are the contours of $\tau_{\rm
    dis}=\big[1,\frac{3}{2},2\big]$ (right to left), as theoretically
  predicted in \cite{pr04}.}
\label{fig:r_T_lin_nlin_ib}
\end{figure}
%%%%%%%%%%%%%%%%%%%%%%%%%%%%%%%%%%%%%%%%%%%%%%%%%%%%%%%%%%%%%%%%%%%%

%%%%%%%%%%% FIGURE 2 %%%%%%%%%%%%%%%%%%%%%%%%%%%%%%%%%%%%%%%%%%%%%
\begin{figure*}[t!]
  \centering
  \includegraphics[width=.45\linewidth]{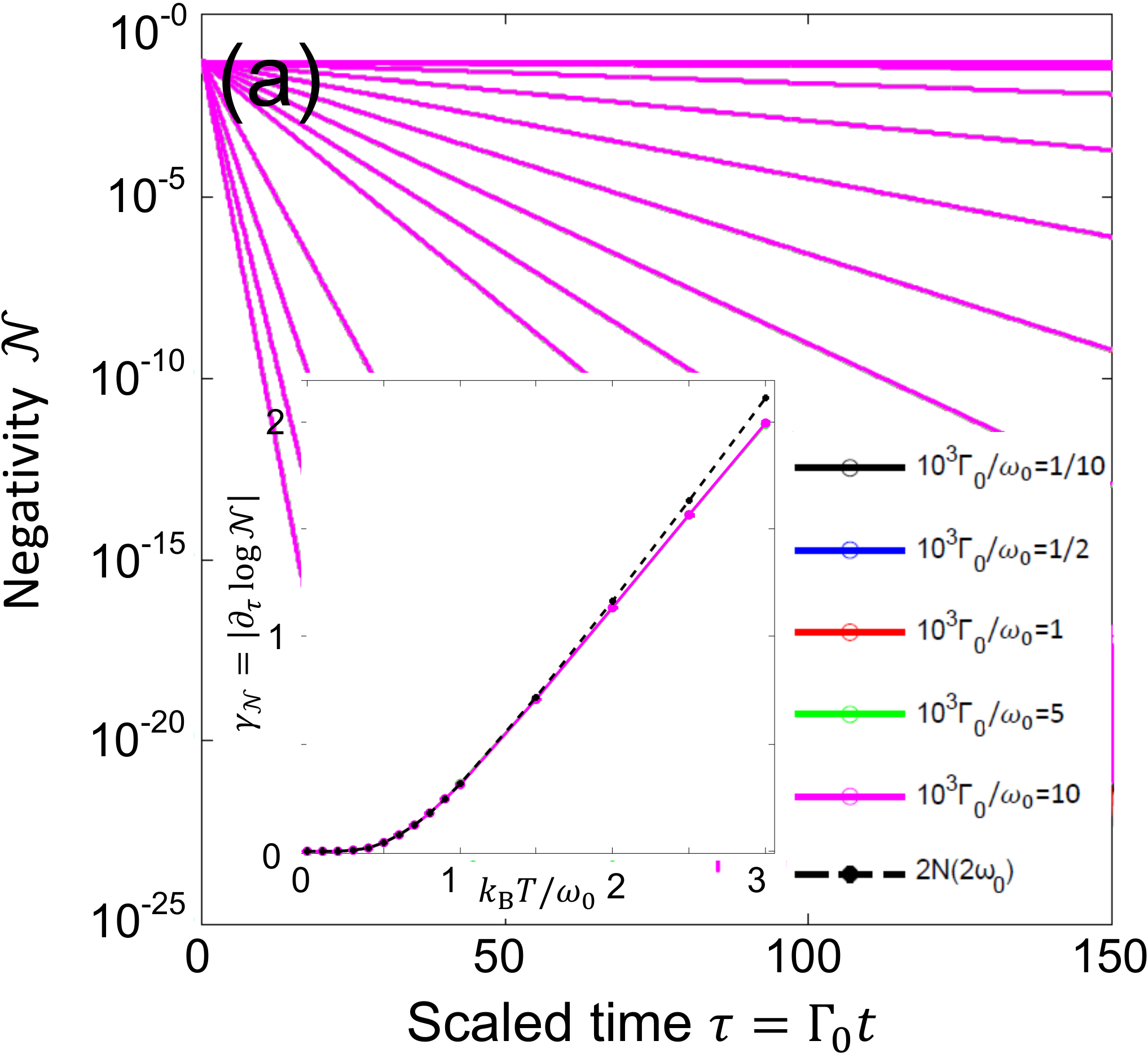}\includegraphics[width=.43\linewidth]{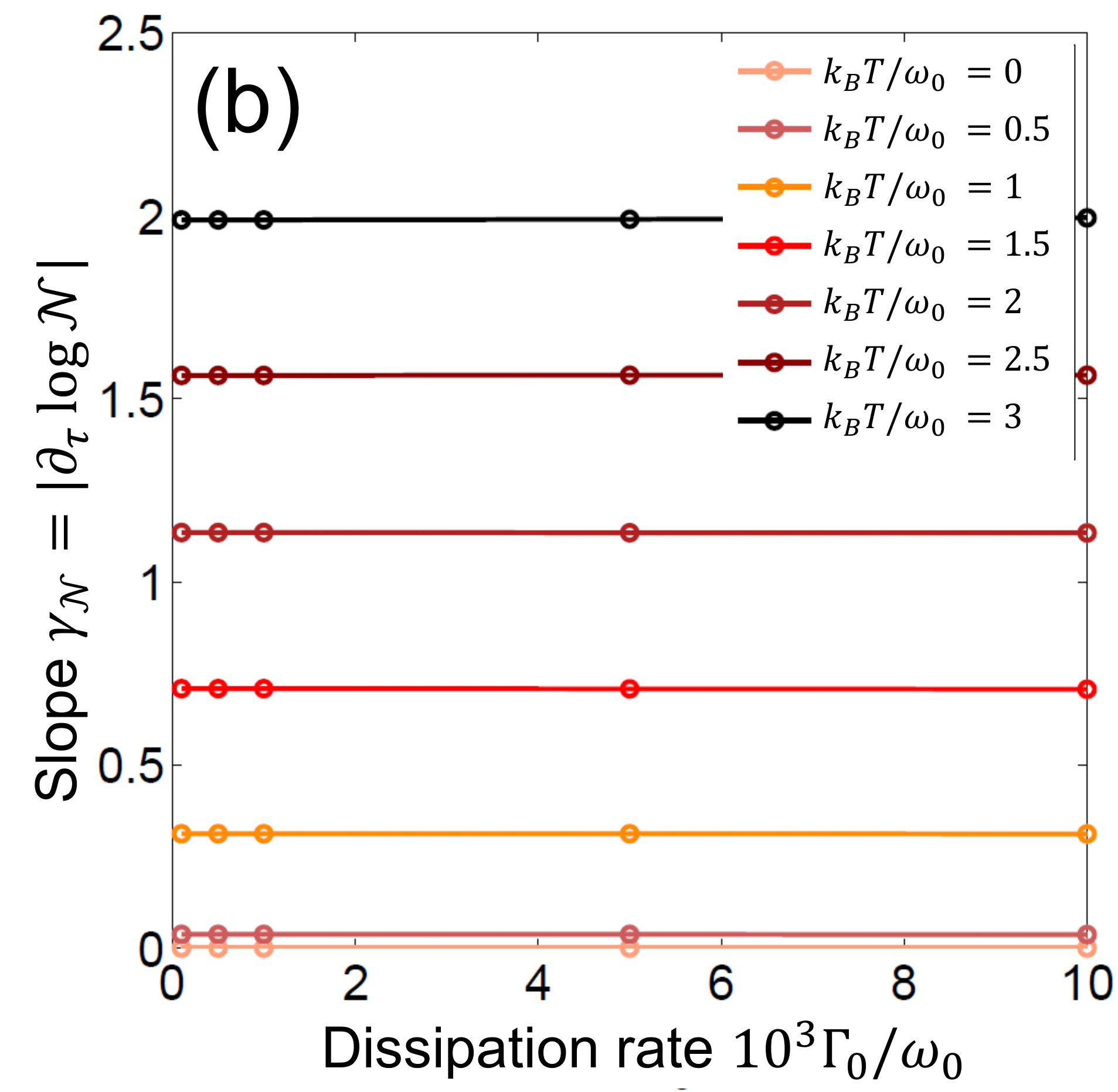}
\caption{ {\bf(a)} Main figure: Time evolution of the negativity of a
  nonlinearly damped squeezed two-mode vacuum with squeezing parameter
  $r=1/20$, individual bath configuration, for temperatures $0\le
  k_{\rm B}T/\omega_0 \le 3$ and damping rates
  $\Gamma_{0}=10^{-3}\omega_0[1/10,1/2,1,5,10]$. The graphs for the
  same $T$ and different $\Gamma_{0}$ overlap when plotted as function
  of $\tau=\Gamma_{0}t$. Inset: Slope of the negativity,
  $\gamma_{\mathcal{N}}=|\partial_\tau \log{\cal N}|$, extracted from
  the second half of the points in the main figure, as function of $T$
  for $\Gamma_{0}=10^{-3}\omega_0[1/10,1/2,1,5,10]$ (color code). The
  dashed line is a slope fit of $2N(2\omega_0)$.  {\bf(b)} Slope of
  the negativity $\gamma_{\mathcal{N}}=|\partial_\tau \log{\cal N}|$
  from the inset in (a) as function of $10^3\Gamma_{0}/\omega_0$.}
\label{fig:vary_gamma_nlin_ib}
\end{figure*}
%%%%%%%%%%%%%%%%%%%%%%%%%%%%%%%%%%%%%%%%%%%%%%%%%%%%%%%%%%%%%%%%%%%%

\subsection{Asymptotic entanglement for coupling to individual baths}\label{sec:ent_ib}
For the individual bath configuration and a Markovian model in the RWA,
it was shown in Ref.~\cite{pr04} that for finite temperatures ($T>0$) all
linearly damped two-mode squeezed vacuum states disentangle
within a finite time, and relax to the ground state. At $T=0$ the
relaxation to the ground state occurs in the limit of infinite time.
These results were further probed with non-Markovian models in
Refs.~\cite{maol+07,ligo07}, lending support to the predictions of
Ref.~\cite{pr04}. Thus, concluding that the Markovian and
non-Markovian dynamics coincide for times larger than reservoir
correlation times.

For NLD, earlier studies of a single oscillator systems
undergoing NLD show that parity conservation brings the system to
a final non-classical steady state~\cite{silo78,vocr+13}. For a
bipartite system with no inter-mode coupling, it follows that the same parity
conservation will, at $T=0$, bring the system into a general steady
state 
\begin{eqnarray}\label{eq:gen_steady_state}
\rho(\infty)&=& P_{00}|00\rangle \langle00 | + P_{11}|11 \rangle
\langle 11 | + \rho_{01,10}|01\rangle \langle 10 | \nonumber \\ & & +
\rho_{10,10}|10\rangle \langle 10 | + \Big[\rho_{00,11}|00\rangle
  \langle 11 | + \nonumber\\ & & \rho_{00,01}|00\rangle \langle 01 | +
  \rho_{00,10}|00\rangle \langle 10 | + \rho_{01,10}|01\rangle \langle
  10 | \nonumber \\ & & +\rho_{01,11}|01\rangle \langle 11 | +
  \rho_{10,11}|10\rangle \langle 11 | + \rm{H.c.} \Big],
\end{eqnarray}
with matrix elements $\rho_{ni,mj}$ determined by the initial state.
The particular initial state \eqref{eq:2modesqvac_fock} leads to a
steady state of the form~\eqref{eq:gen_steady_state} where several
elements are zero, reducing it to
\begin{multline}\label{eq:sqvac_steady_state}
\rho(\infty)= P_{00}|00\rangle \langle00 | + 
P_{11}|11 \rangle \langle 11 | \\ + 
(\rho_{00,11}|00\rangle \langle 11 | + \rm{H.c} )\;.
\end{multline}

The element $\rho_{00,11}$ is important for the
asymptotic negativity (entanglement). While initially there are
multiple off-diagonal elements contributing to the negativity
$\mathcal{N}(t=0)=(e^{2|r|}-1)/2$~\cite{paro08}, these quickly
decohere, leaving only the parity protected matrix elements in
(\ref{eq:sqvac_steady_state}), and the negativity saturates at
$\mathcal{N}(\infty)=|\rho_{00,11}(\infty)|_{T=0}$. This can be
verified through the characteristic equation for $\rho^{{\rm T}_1}$ of
the steady state \eqref{eq:sqvac_steady_state}. For a general
$M\times M$ basis size the characteristic equation is given by
\begin{equation}
(-\mu)^{M-4}(P_{00}-\mu)(P_{11}-\mu)(\mu^2-|\rho_{00,11}|^2)=0,
\end{equation}
with only one negative root $\mu=-|\rho_{00,11}|$.

Comparing the nonlinear and linear decays of squeezed states at $T=0$ 
one finds that the nonlinearly damped states remain
entangled with a saturating negativity, whereas the linearly damped
states asymptotically disentangle in the limit of $t\rightarrow
\infty$. This can be seen in Fig.~\ref{fig:r_T_lin_nlin_ib} showing
the scaled disentanglement time $\tau_{\rm dis}$ (colorbar) of a
nonlinearly damped (main figure) and linearly damped (inset) two-mode
squeezed vacuum as function of $T$ and $r$. The results are obtained from
numerical simulations. The scaled disentanglement time is defined as
$\tau_{\rm dis}=\Gamma_0 t_{\rm{dis}}$, where $t_{{\rm dis}}$ is the
time at which $\mathcal{N}<\varepsilon$, and $\varepsilon$ is the negativity
cut-off. Temperature is measured in units of $\hbar\omega_0/k_{\rm B}$. The main
figure and inset have identical simulation parameters, but different
disentanglement time scales (main and inset colorbars). The white,
dashed lines in the inset are the theoretically predicted
disentanglement times derived in Ref.\ \cite{pr04}.

The nonlinearly and the linearly damped systems for
$T>0$ both display a finite disentanglement time for all
$r$. The main difference is the disentanglement time scale. The
nonlinearly damped states disentangle much slower than the linearly
damped states. During the chosen evolution time all LD states
disentangle, while a part of the NLD states remain entangled (white
region, main figure). After a longer time evolution all NLD states
will eventually disentangle.

The time evolution of the negativity while relaxing to the steady state
is shown in Fig.~\ref{fig:vary_gamma_nlin_ib} (a). Here we
use a constant value of $r=1/20$ for several values of $T$ and
$\Gamma_{0}$. The graphs for the same $T$ and different $\Gamma_{0}$
overlap when expressed in terms of scaled time units
$\tau=\Gamma_{0}t$. Initially the negativity has a rapid initial
transient during which the initial squeezed state reduces to a state
of the form given by Eq.~(\ref{eq:sqvac_steady_state}). This is
followed by a slow exponential decay.

To quantify the asymptotic decay, the inset in
Fig.~\ref{fig:vary_gamma_nlin_ib}(a) shows the negativity slope
$\gamma_{\mathcal{N}}=|\partial_{\tau} \log {\cal N}|$, extracted
from the second half of the data in the main panel of
Fig.~\ref{fig:vary_gamma_nlin_ib}(a), as function of $T$. The decay
rate solely depends on $T$, which is further corroborated in
Fig.~\ref{fig:vary_gamma_nlin_ib} (b), showing the negativity slope
$\gamma_{\mathcal{N}}$ from (a) as function of $10^{3}\Gamma_{0}$.
As shown in appendix~\ref{App:AppendixA}, in the limit of low
temperatures, within the RWA, the slope is given by the expression
$\mathcal{N}(\tau)=(|\rho_{00,11}(\infty)|_{T=0})e^{-2N(2\omega_0)\tau}$,
shown as the dashed line in the inset to
Fig.~\ref{fig:vary_gamma_nlin_ib}(a).

The much slower decay of the negativity in the nonlinearly damped case
compared with the linearly damped system can be understood as follows:
For finite temperatures the disentanglement of the nonlinearly damped
squeezed vacuum states is related to the slow, thermal dephasing of
the parity protected matrix element $\rho_{00,11}$. However, thermal
decoherence of this element requires a simultaneous two-quanta
excitation of both oscillators, a process which is less probable
compared to linear decoherence, where neither an individual nor
simultaneous excitation is needed to achieve a de-excitation.
%%%%%%%%%%% FIGURE 3 %%%%%%%%%%%%%%%%%%%%%%%%%%%%%%%%%%%%%%%%%%%%%
\begin{figure}[t]
  \centering
\includegraphics[width=\linewidth]{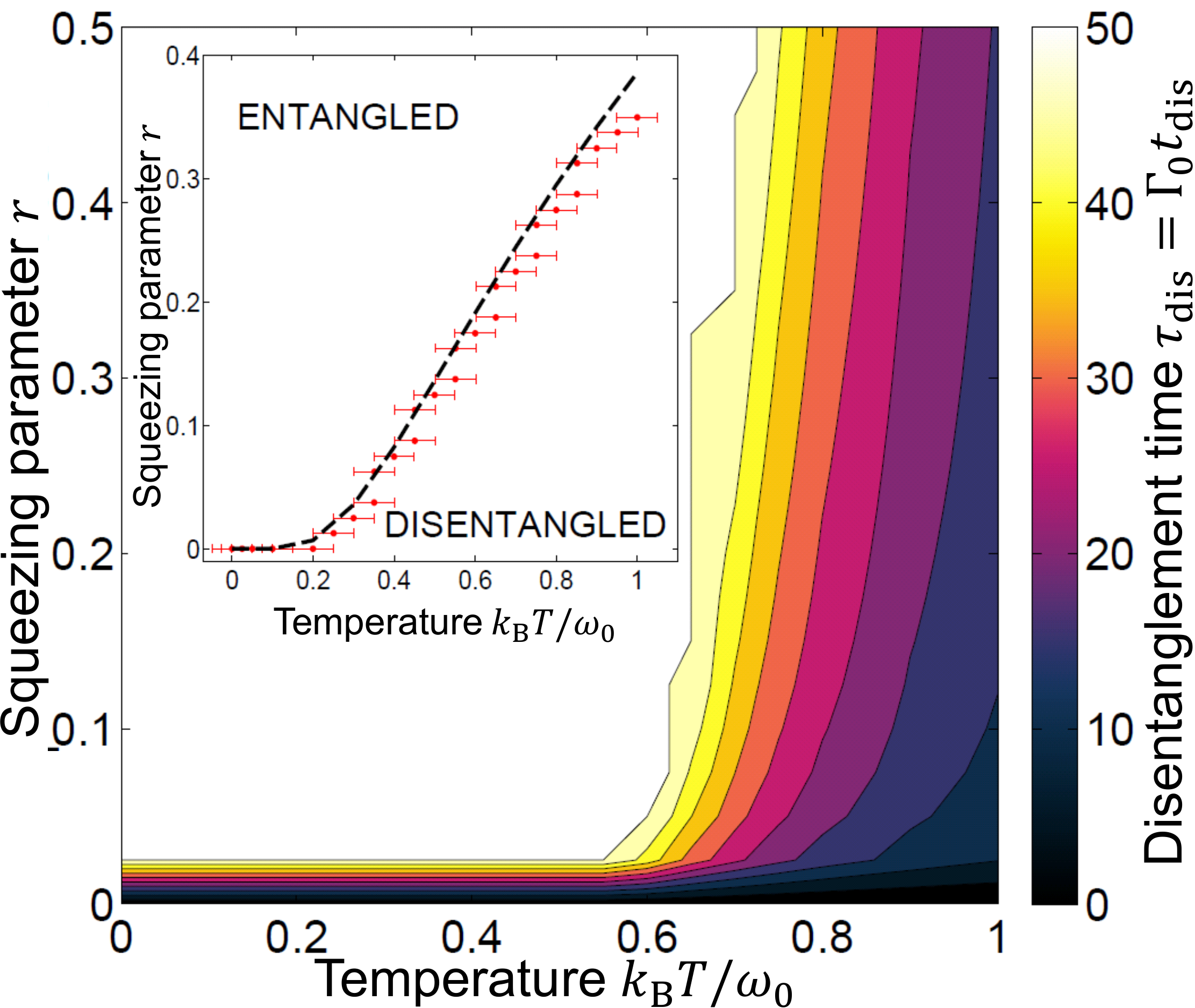}
\caption{Main figure: disentanglement time $\tau_{\rm dis}=\Gamma_0
  t_{\rm dis}$ (colorbar) of nonlinearly damped two-mode squeezed
  vacuum states as function of temperature $T$ and squeezing parameter
  $r$, common bath, with simulation time $\tau_{\rm sim}=50$, damping
  rate $\Gamma_{0}=10^{-3}\omega_0$, and the negativity cut-off
  $\varepsilon=10^{-3}$. Inset: Entanglement borderline of the
  linearly damped two-mode squeezed vacuum states, common bath, in the
  phase space of $r$ and $T$. The dashed line is the theoretical
  prediction in \cite{pr04} and the dots with errorbars are numerical
  data. The simulation parameters are: $\tau_{\rm sim}=50$,
  $\Gamma_{0}=10^{-3}\omega_0$ and $\varepsilon=10^{-3}$.
}\label{fig:r_T_lin_nlin_cb}
\end{figure}
%%%%%%%%%%%%%%%%%%%%%%%%%%%%%%%%%%%%%%%%%%%%%%%%%%%%%%%%%%%%%%%%%%%%%%

%%%%%%%%%%% FIGURE 4 %%%%%%%%%%%%%%%%%%%%%%%%%%%%%%%%%%%%%%%%%%%%%
\begin{figure}[t!]
  \centering
\includegraphics[width=\linewidth]{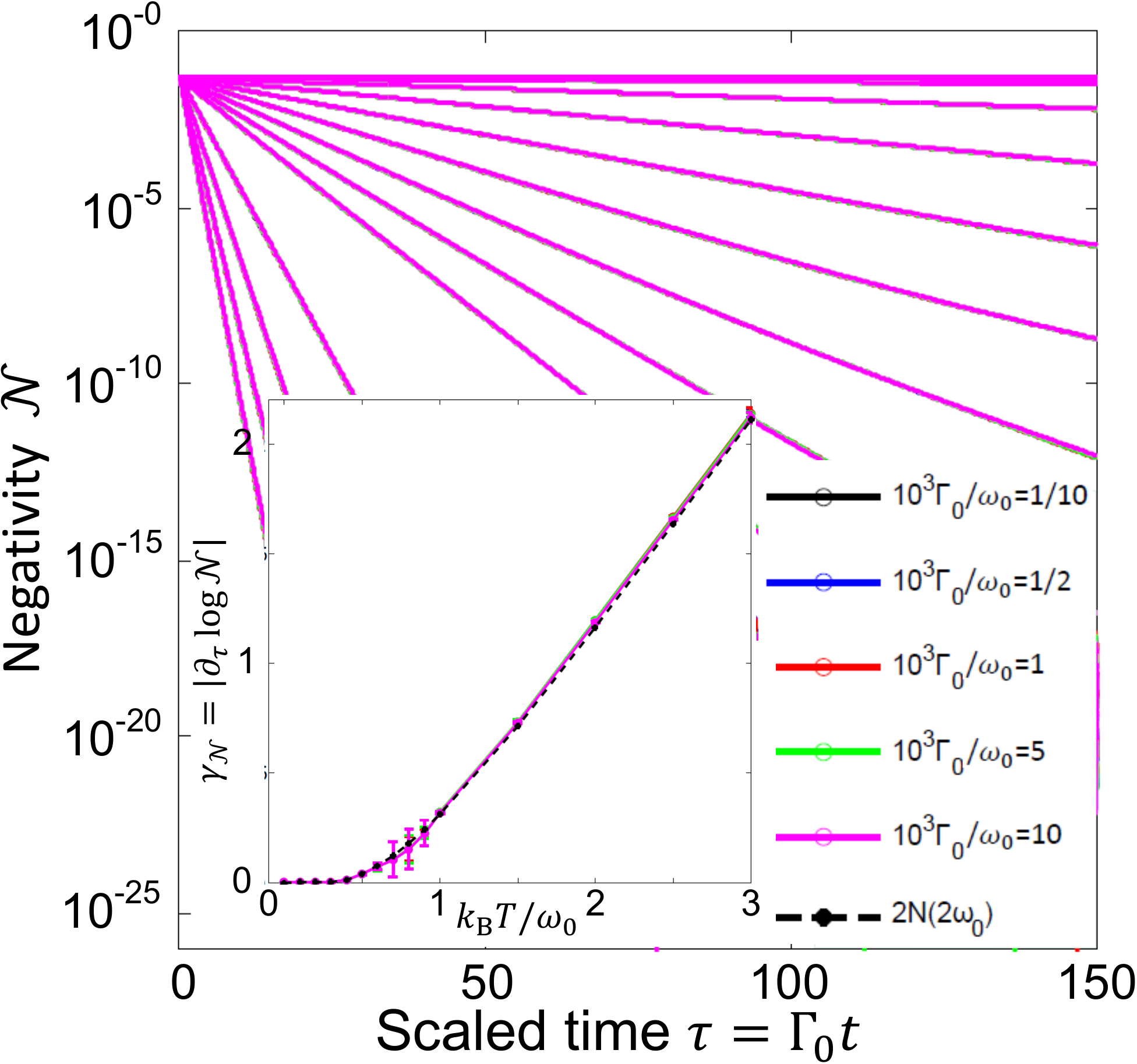}
\caption{ (Color online) Time evolution of the negativity of a
  nonlinearly damped squeezed two-mode vacuum with squeezing parameter
  $r=1/20$, common bath configuration, for temperatures $0\le k_{\rm
    B}T/\omega_0 \le 3$ and damping rates
  $\Gamma_{0}=10^{-3}\omega_0[1/10,1/2,1,5,10]$. The inset shows the
  slope $\gamma_{\mathcal{N}}$ extracted from the second half of the points 
  in the main figure, as
  function of $T$ for $\Gamma_{0}=10^{-3}\omega_0[1/10,1/2,1,5,10]$
  (color code). The dashed line is the function $2N(2\omega_0)$. }
\label{fig:vary_gamma_nlin_cb}
\end{figure}
%%%%%%%%%%%%%%%%%%%%%%%%%%%%%%%%%%%%%%%%%%%%%%%%%%%%%%%%%%%%%%%%%%%%%

\subsection{Asymptotic entanglement for coupling to a common bath}
We now turn to the situation with the two independent oscillators
nonlinearly coupled to a common reservoir. For a linearly damped
system in this configuration, the asymptotic steady states of an
initial two mode squeezed state (\ref{eq:2modesqvac}) can be divided
into entangled and separable states~\cite{pr04,paro08,paro09}. To
which category a state will belong, depends on $r$ and $T$. This
result was obtained in Ref.\ \cite{pr04} using Markovian dynamics and
RWA and is shown in Fig.\ \ref{fig:r_T_lin_nlin_cb}(inset). An
interesting feature is that the system never disentangles for
$T=0$. This entanglement preservation can be explained in terms of
normal modes, e.g.\ center of mass and relative coordinates. As only
the center of mass motion is affected by the dissipation to the bath,
the relative motion of the oscillators evolves freely. For nonlinear
system reservoir coupling, there is no decoupling of the relative
oscillator motion from the bath, and we do not find any finite
temperature steady state entanglement.

As seen in the main figure \ref{fig:r_T_lin_nlin_cb}, showing the
scaled disentanglement time (colorbar) of nonlinearly damped two-mode
squeezed vacuum states as function of $T$ and $r$ for the common bath
configuration, the asymptotic entanglement behavior is very similar to
the nonlinearly damped squeezed states in the individual bath
configuration.  Again, the nonlinearly damped states disentangle
slower than the linearly damped states in the disentangled region in
the inset of Fig.~\ref{fig:r_T_lin_nlin_cb}.  Like in the main panel
of Fig.~\ref{fig:r_T_lin_nlin_ib}, not all states have yet
disentangled for the chosen simulation time (white region), but will
do so after a longer evolution.

The similarity in the behaviors stem from a supression of information
exhange between the oscillators via the bath in the steady state
$\rho(\infty)$ for $T\approx0$. In the QME
(\ref{eq:QME_RWA_wIntfinal_cb}) the term $\mathcal{L}_1\rho$ quickly
brings the initial state (\ref{eq:2modesqvac}) to the steady state
(\ref{eq:sqvac_steady_state}), which cannot be further affected by the
term $\mathcal{L}_2\rho$, as $\mathcal{L}_2\rho(\infty)=0$. For higher
temperatures, the term $\mathcal{L}_2\rho$ will contribute to some
information exchange, but not enough to significantly alter the
influence of the $\mathcal{L}_1\rho$-term. The qualitative evolution
of the state is therefore as for the individual bath
configuration. This is supported by the results in
Fig.~\ref{fig:vary_gamma_nlin_cb}, displaying the slow temperature
dependent negativity decay for various $T$ and $\Gamma_0$ (main
figure), and the temperature dependent negativity decay-rates
$\gamma_{\cal N}=|\partial_\tau \log {\cal N}| = 2N(2\omega_0)$
(inset). The derivation of the exponent can be found in appendix
\ref{App:AppendixA} and is equal to the thermal decay exponent
obtained for individual baths.

\section{Results for coupled oscillators\label{sec:coupled}}
Finally, we consider two weakly coupled oscillators. Based on the
results of the preceding section, the main qualitative difference
between LD and NLD stems from the parity conservation for the
individual oscillators. When the oscillators are coupled only the
parity of the entire system is conserved. In particular, the element
$\rho_{00,11}$ which we found to give the asymptotic negativity, is no
longer protected and can decay via an intermediate transition to the
state $\rho_{00,20}$.

Since the situation is close to the linear case, we restrict the
discussion to individual baths. The system Hamiltonian
(\ref{eq:hamiltonian_terms1}) is adjusted to include an inter-mode
coupling term
\be\label{eq:HScoupling}
H_{\rm S} = \sum_{j=1,2}
\left(\frac{1}{2}p_j^2+\frac{1}{2}\omega_j^2 q_j^2 \right)
+ \sqrt{\omega_1 \omega_2}\lambda q_1 q_2 \;.
\ee
The corresponding Hamiltonian in the RWA is
\be\label{eq:HScouplingRWA}
H_{\rm S,RWA} = \sum_{j=1,2} \omega_j a^{\dag}_j a_j
+ \frac{\lambda}{2}a^{\dag}_j a_{j-(-1)^j}\;.
\ee
By the same procedure as described in
Sec.\ \ref{sec:model}, the RWA-QME for a symmetric system,
$\omega_j=\omega_0$, in the weak intermode coupling limit,
$\lambda\ll\omega_0$, obtains the form \cite{vois+13}
\begin{eqnarray}\label{eq:QME_RWA_Coupled_ib}
\dot{\rho} &= & -\frac{1}{2}\sum_{j=1,2}
\Big[\gamma_j(2\omega_0)\mathcal{L}_1[a_j^{\dag 2}] + \\\nonumber & &
  \gamma_j(-2\omega_0)\mathcal{L}_1[a_j^2] \Big]\rho
+\mathcal{D}_{12}(\lambda)\rho\;.
\end{eqnarray}
To the lowest order in $\lambda$ the oscillators are individually
coupled to their respective reservoirs, and the superoperator
$\mathcal{D}_{12}(\lambda)$ becomes
\begin{align}
	\mathcal{D}_{12}(\lambda)\rho
	={}& \Upsilon_{+} \mathcal{L}_1[(n_1 - n_2)]\rho
        \notag\\ {}& -\frac{1}{2}\Upsilon_{-} \Big[ (n_1 - n_2)
          (a^\dag_1 a_2 - a^\dag_2 a_1) \rho \Big.\notag\\ {}&-
          (a^\dag_1 a_2 - a^\dag_2 a_1) \rho (n_1 - n_2)
          \notag\\ {}&+ \rho (a^\dag_1 a_2 - a^\dag_2 a_1)^\dag
          (n_1 - n_2)\notag\\ {}&\Big.- (n_1 - n_2) \rho (a^\dag_1
          a_2 - a^\dag_2 a_1)^\dag \Big]\;. \label{eq:cross_D}
\end{align}
Here $\Upsilon_{\pm} = \gamma(\lambda)\pm\gamma(-\lambda)$ with
$\gamma(\lambda) = \kappa(\lambda) [N(\lambda) + 1]$ and
$\gamma(-\lambda) = \kappa(\lambda) N(\lambda)$.

The coupling $\lambda$ plays a dual role of contributing to oscillator
interaction via $H_{\rm S}$ and to decoherence via $\mathcal{D}_{12}$.
For linearly damped oscillators the decoherence terms in
$\mathcal{D}_{12}$ would only arise if $\Gamma_1 \neq \Gamma_2$
\cite{pool+05}. There is no such restriction however for the
nonlinearly damped system where the superoperator $\mathcal{D}_{12}$
consists of two decoherence terms, proportional to $\Upsilon_+$ and
$\Upsilon_-$ respectively. The temperature dependencies of these terms
differ from those of $\gamma_j$. For temperatures exceeding the
coupling energy, $\lambda\ll k_{\rm B} T$, we have $\Upsilon_{+} >
\Upsilon_{-}$. In the low temperature limit, $\lambda\gg k_{\rm B} T$,
both terms approach equal magnitudes $\Upsilon_{+} \approx
\Upsilon_{-}$.

%%%%%%%%%%% FIGURE 5 %%%%%%%%%%%%%%%%%%%%%%%%%%%%%%%%%%%%%%%%%%%%%
\begin{figure}[t]
  \centering \includegraphics[width=\linewidth] {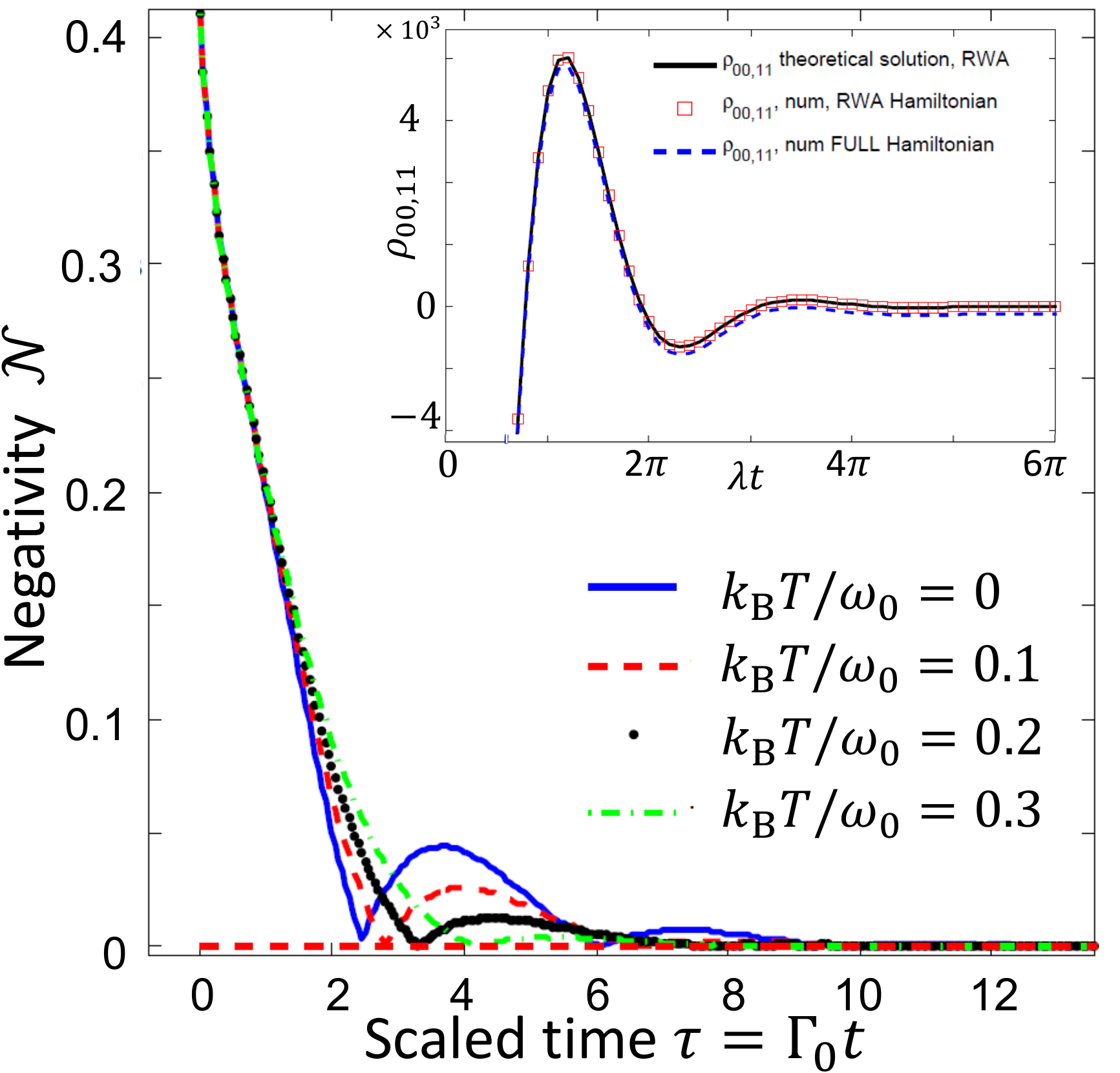}
\caption{Decay of the negativity $\cal N$ as function of scaled time
  $\tau=\Gamma_0 t$ for an initial state with squeezing parameter
  $r=0.3$ for different temperatures and
  $\lambda=\Gamma_{0}=10^{-3}\omega_0$. The coupling is reflected in
  the coherent oscillations superposed on the exponential decay
  towards the thermal equilibrium state. The inset shows the time evolution of
  $\rho_{00,11}$ of a nonlinearly damped squeezed two-mode vacuum with
  squeezing parameter $r=1/20$, individual bath configuration, for
  $T=0$ and damping rate $\lambda=\Gamma_{0}=10^{-3}\omega_0$. }
\label{fig:fig5}
\end{figure}
%%%%%%%%%%%%%%%%%%%%%%%%%%%%%%%%%%%%%%%%%%%%%%%%%%%%%%%%%%%%%%%%%%%%%%%%

For weakly coupled oscillators, linearly coupled to individual baths,
a phase diagram separating the entangled steady states from
non-entangled steady states exists (see for instance
Ref.~\cite{gapa+10}).  Whether or not the state remains entangled
depends on temperature and strength of coupling. A more in depth
study, using both Markovian as well as non-Markovian evolution can be
found in Ref.~\cite{ligo07}. As explained in Ref.~\cite{ligo07},
starting from an initial two-mode squeezed state, an undamped coupled
oscillator system will display coherent oscillations during its time
evolution with corresponding oscillations in negativity. Adding finite
linear damping, results in loss of entanglement in the long time limit
and suppression of coherent oscillations in the negativity.

For the case of nonlinear coupling to individual baths we find results
which are similar to those in Ref.~\cite{ligo07}, with coherent
oscillations reflected in the decay curve shown in
Fig.~\ref{fig:fig5}. As can be seen, with increasing temperature the
oscillations vanish but the initial rapid decay remains unaltered. 
The coherent oscillations can be traced back to the time evolution for
$T=0$ and $\lambda>0$. In this case, after the transient rapid decay,
the negativity dynamics is governed by the evolution of the
$\rho_{00,11}$ matrix element. Since the two-mode squeezed vacuum only
has even entries, it suffices to analyze the matrix elements with
total amount of two quanta. The evolution of $\rho_{00,11}$ is
influenced by the elements $\rho_{00,02}$ and $\rho_{00,20}$ which in
the RWA approximation contribute to decoherence and hence the
asymptotic negativity decays to zero. The detailed calculation is
given in Appendix \ref{app:ib_coupling}. The resulting time evolution
of $\rho_{00,11}$ is shown in the inset of Fig.~\ref{fig:fig5}.

Numerically solving the full QME at $T=0$, we see a residual nonzero
negativity. This is due to the RWA-Hamiltonian ($H_{\rm S}\propto H_0
+ \lambda(a^{\dag}_1 a_2 + a^{\dag}_2 a_1)$) not properly reproducing
the correct ground state of the coupled oscillator system since terms
proportional to $a^2$ ($a^\dagger a^\dagger$) are neglected. Hence,
the numerical results display a small residual entanglement. In the
inset of Fig.~\ref{fig:fig5} the comparison of the results of the
numerical simulation and the RWA shows a very good agreement.

\section{Conclusions}
We have studied the asymptotic behavior of entanglement between two
harmonic oscillators when they are quadratically coupled to an
environment. In particular, we have investigated to what extent
phenomena, known from studying the decay of two-mode squeezed states in
the corresponding linearly damped systems, change when damping is
nonlinear. We find that the number parity conservation associated with
pure nonlinear damping causes significant reduction of the
disentanglement rate. Moreover, the equilibrium distribution is
different from the standard Bose-distribution. Further, in contrast to
the linearly damped systems, we find no qualitative difference between
oscillators coupled to common baths and coupled to individual
baths. We attribute the latter effect to the lack of a conserved
quantity (relative oscillator energy) in the nonlinearly coupled
system in compination with a suppressed information exhange between
the oscillators at low temperatures. For weakly coupled oscillators,
the number parity is no longer individually conserved, hence, the
system can relax to the ground state.

The results here are obtained in the Markovian limit. Extending the
study to a non-Markovian dynamics could alter the picture presented
here. For instance, it is known that for linearly damped oscillator
systems studied with non-Markovian dissipation models~\cite{ligo07,
  paro08, paro09}, a more detailed and complex picture of the
asymptotic behavior emerges.  Still, in those studies the overall
characteristics obtained in the Markovian limit, for instance the
division into entangled and separable steady states, remain intact.

At present it is not known whether it is possible to realize a system
where the dominant dissipation mechanism at low excitation levels is
purely nonlinear. However, for some dissipation mechanisms, see for
instance Ref.~\cite{crmi+12}, symmetry can dictate that the lowest
order coupling to the environment must be quadratic in the
coordinates.  Systems with such symmetries are thus strong candidates
for studying NLD in the quantum regime. Moreover, it was suggested
that engineering of NLD might be feasible
\cite{nubo+10,lemi04,evsp+14}.  Exploiting the reduced disentanglement
in systems with NLD is a promising path towards realizing entanglement
based technologies.

\acknowledgments
Financial support for this work was provided by the Swedish Research Council VR. 

\appendix
\section{}\label{App:AppendixA}

\subsection{Negativity exponent - individual baths.}\label{app:ib_exponent}
Here it is shown that $\gamma_{\cal N}=2N(2\omega_0)$.
As argued in Sec.\ \ref{sec:ent_ib}, $\mathcal{N}(\tau)$ of a nonlinearly damped
squeezed state asymptotically only depends
on the density matrix element $\rho_{00,11}$. 
Equation (\ref{eq:QME_RWA_wIntfinal_cb}) is used to obtain the
respective equation of motion, along with equations of
motion for elements $X=\sqrt{3}(\rho_{02,13}+\rho_{20,31})$,
which influence $\rho_{00,11}$.
By assuming that the other elements have
already decayed, one obtains a set of 
two coupled, first order differential equations
\begin{subequations}\label{eq:diff_eqs_nlin_ib}
\begin{align}
\dot{\rho}_{00,11}&=2\gamma_{2-}X-8\gamma_{2+}\rho_{00,11}, \label{eq:psi00}\\
\dot{X}&=12\gamma_{2+}\rho_{00,11} -4aX \label{eq:X},
\end{align}
\end{subequations}
where $a=(\gamma_{2-}+5\gamma_{2+})$ and $\gamma_{2\pm}$ is scaled by $\Gamma_0$.
For solutions of the form of $\rho_{00,11}\sim Ae^{r_1\tau}+Be^{r_2\tau}$ on finds, 
\be\label{eq:charRoots}
r_{1,2}=-(16\gamma_{2+}+2)\pm 2[(8\gamma_{2+}+1)^2-2\gamma_{2+}(21\gamma_{2+}+1)]^{\frac{1}{2}}\;,
\ee
where $\gamma_{2-}=\gamma_{2+}+1$ was used.
For low $T$, $\gamma_{2+}=N(2\omega_0)<1$, the square root in
(\ref{eq:charRoots}) can 
be Taylor expanded up to second order to yield
\be
r_1 \approx -2N(2\omega_0), \:\: r_2\approx -2(15N(2\omega_0) +2),
\ee
and
\be
\rho_{00,11}(\tau)= Ae^{-2N(2\omega_0)\tau}+Be^{-2(15N(2\omega_0)+2)\tau},
\ee
where the first term denotes the slow thermal decay. The second term
describes a rapid initial decay of the matrix element.
The amplitudes are given by
\begin{subequations}
\begin{align}
A&=\frac{\rho_{00,11}(0)(11N(2\omega_0)+2)}{14N(2\omega_0)+2},\\
B&=\rho_{00,11}(0)\Big[1-\frac{11N(2\omega_0)+2}{14N(2\omega_0)+2}\Big],
\end{align}
\end{subequations}
where $\rho_{00,11}(0)$ is given by (\ref{eq:2modesqvac_fock}).
Also, $A>0$ for all $r>0$ and $A\gg B$ for low $T$. Hence $\gamma_{\cal N}=2N(2\omega_0)\tau$.

\subsection{Negativity exponent - common bath.}\label{app:cb_exponent}
Here we verify that the decay of entanglement, as governed by the matrix element
$\rho_{00,11}$, for the individual bath case is $\gamma_{\cal N}=2N(2\omega_0)$.
With the same assumptions as for the individual baths, 
from equation (\ref{eq:QME_RWA_wIntfinal_cb})
one obtains the equations of motion for the
matrix elements responsible for the negativity in the
common bath case
\begin{subequations}
\begin{align}
\dot{\rho}_{00,11}&=2\sqrt{3}\gamma_{2-}Z-8\gamma_{2+}\rho_{00,11},\\
\dot{Z}&= 8\sqrt{3}\gamma_{2+}\rho_{00,11}-8(\gamma_{2-}+3\gamma_{2+})Z, 
\end{align}
\end{subequations}
where $Z=\rho_{02,13}+\rho_{02,31}+\rho_{20,13}+\rho_{20,31}$. In this
case we obtain the solution
\be
\rho_{00,11}= C e^{-2N(2\omega_0)\tau}+D e^{-2(19N(2\omega_0)+4)\tau},
\ee
with the amplitudes 
\begin{subequations}
\begin{align}
C&=\frac{\rho_{00,11}(0)(15N(2\omega_0)+4)}{18N(2\omega_0)+4}\\ D&=\rho_{00,11}(0)\Big[1-\frac{15N(2\omega_0)+4}{18N(2\omega_0)+4}\Big],
\end{align}
\end{subequations}
where $\rho_{00,11}(\tau=0)$ is given in (\ref{eq:2modesqvac_fock}).
Also, $C>0$ for all $r>0$ and $C\gg D$ for low $T$. Hence the decay rate is again
dominated by $\gamma_{\cal N}=2N(2\omega_0)$.\\
 
\subsection{Negativity evolution - individual bath, non-zero inter-mode coupling.}\label{app:ib_coupling}
For $T=0$ and an inter-mode coupling $\lambda>0$ the evolution of
negativity is still governed by the $\rho_{00,11}$ element. For an
initially two-mode squeezed state, the equations of motion governing
the evolution are
\begin{subequations}\label{eq:diffeqs_ib_coupling_T0}
\begin{align}
\dot{\rho}_{00,11}&=\frac{\rm{i} \lambda}{\sqrt{2}} Y\;,\\
\dot{Y}&=(\rm{i}\sqrt{2}\lambda-2\sqrt{2}\Upsilon)\rho_{00,11}
-(\gamma_{2-}+2\Upsilon)Y,
\end{align}
\end{subequations}
where $Y=\rho_{00,02} + \rho_{00,20}$. For a solution of the form
$\rho_{00,11}=A_{+}e^{r_{+}\tau} + A_{-}e^{r_{-}\tau}$
one finds
\be
r_{\pm}=-\frac{1}{2}(\Gamma_0 +2\Upsilon) 
\pm \frac{1}{2}\sqrt{(\Gamma_0 +2\Upsilon)^2-4(\lambda^2 +\rm{i}2\lambda\Upsilon)},
\ee
with amplitudes
\begin{subequations}
\begin{align}
A_{+} &= -\frac{\rho_{00,11}(0) r_{-}}{(r_{+}-r_{-})}\;,\\
A_{-} &=\rho_{00,11}(0)\Big[1+\frac{r_{-}}{(r_{+}-r_{-})}\Big]\;.
\end{align}
\end{subequations}
\bibliography{edqstb}
\end{document}